\documentstyle[epsf,prl,twocolumn,aps]{revtex}
\tighten 

\begin{document}
\draft
\title{Spin splitting in a polarized quasi-two-dimensional exciton gas}
\author{L. Vi\~na, L. Mu\~noz and E. P\'erez}
\address{Departamento de F\'{\i}sica de Materiales, C-IV\\
Universidad Aut\'onoma de Madrid, Cantoblanco, E-28049, Madrid, Spain.}
\author{J. Fern\'{a}ndez-Rossier and C. Tejedor}
\address{Departamento de F\'{\i}sica Te\'orica de la Materia Condensada,\\
Universidad Aut\'onoma de Madrid, Cantoblanco, E-28049, Madrid, Spain.}
\author{K. Ploog}
\address{Paul Drude Institut fuer Festkoerperelektronik,\\
Hausvogteiplatz 7, Berlin, D-10117, Germany.}
\date{\today}

\twocolumn[

\maketitle  \widetext   \vspace*{-1.0truecm}   

\begin{abstract}
\begin{center} 
\parbox{14cm}{We have observed a large spin splitting between "spin" $+1$ 
and $-1$ heavy-hole excitons, having unbalanced populations, in undoped 
GaAs/AlAs quantum wells in the absence of any external magnetic field.
Time-resolved photoluminescence spectroscopy, under excitation with
circularly polarized light, reveals that, for high excitonic
density and short times after the pulsed excitation,
the emission from majority excitons lies above that of minority ones. 
The amount of the splitting, which can be as
large as 50$\%$ of the binding energy, increases with excitonic density and
presents a time evolution closely connected with
the degree of polarization of the luminescence.
Our results are interpreted on the light of a recently 
developed model, which shows that,
while intra-excitonic exchange interaction is
responsible for the spin relaxation processes, exciton-exciton interaction  
produces a breaking of the spin degeneracy in two-dimensional
semiconductors.}
\end{center}
\end{abstract}
\pacs{ \hspace{1.9cm} 
} ] \narrowtext

The spin dynamics of low-dimensional semiconductor heterostructures has been
studied intensively both experimentally\cite
{Damen_91,Stark_92,Dareys_93,Amand_94,Vinattieri_93,Munoz_95} and
theoretically\cite{Bastard_92,Maialle_93,Joaq_96} in the last decade. One of
the most intriguing findings, first observed by Damen et al.\cite{Damen_91}
in time-resolved photoluminescence (TRPL) spectra, is the appearance of an
energy splitting between excitons with spin +1 and -1, \cite{com_1} 
in the absence of any magnetic field, whenever circularly polarized 
light is used to excite the system. The
breaking of the spin degeneracy at high excitonic densities was confirmed by
pump and probe experiments,\cite{Stark_92} and later on by TRPL. \cite
{Dareys_93,Amand_94,Munoz_95b} Closely related to this behavior of
spin-polarized excitons is the existence of a blue shift of excitonic
transitions, observed in pump and probe experiments in GaAs quantum wells
(QWs) under high excitation.\cite{Peyghambarian_84,Hulin_86} This shift
has been attributed to the repulsive interaction among excitons due to the
Pauli exclusion principle acting on the Fermi particles forming the excitons.
\cite{Rink_85,Haug_94} In a recent publication, \cite{Joaq_96} 
we have shown that, in the case of spin dependent populations, 
inter-excitonic interaction produces a breaking of the spin
degeneracy in two-dimensional excitons. This mechanism is
complementary to the intra-exciton exchange, which induces the spin
relaxation. \cite{Maialle_93}

We present in this work new experimental results on spin splitting
of heavy-hole (hh) excitons 
in intrinsic GaAs quantum wells and compare the energies of the
interacting-excitons system with theoretical calculations based on the model
described in Ref. \cite{Joaq_96}. We have used a sample consisting
of 50 periods of 77$\AA$-wide GaAs wells and 72$\AA$-wide AlAs
barriers, which presents a small Stokes
shift between the emission and the absorption (2.5 meV at 2 K and very low
exciting power). The Stokes shift allows us to perform 
quasi-resonant excitation
experiments, i.e. detect in the peak of the photoluminescence while exciting
in the ground-state absorption peak. In spite of the presence of this 
shift, the sample exhibits dynamical properties comparable to those of very
high quality samples. \cite{Munoz_95}

TRPL spectra have been measured with a standard up-conversion setup, using a
double-monochromator to disperse the up-converted signal. The exciting
light, obtained from a Styryl 8 dye laser synchronously pumped by the 532 nm
line of a mode-locked Nd:YAG laser, was circularly polarized by means 
of a $\lambda$/4 plate, and the PL was analyzed into its $\sigma ^{+}$ and $\sigma
^{-}$ components using a second $\lambda$/4 plate
before the non-linear crystal, where the up-conversion taks place. 
Spectra were taken at different times delays with the sample
mounted in the cold finger of a temperature-variable cryostat. The time
resolution of our system is 5 ps.

Figure 1 depicts time-resolved PL spectra at 8 K taken 10 ps after the excitation
with $\sigma ^{+}$ pulses for two different densities, $4 \times 10^{10} cm^{-2}$
and $1.5 \times 10^{11} cm^{-2}$, with the laser at 1.625 eV. 
The solid points
show the polarized ($\sigma ^{+}$, spin +1) emission 
while the open circles
correspond to the unpolarized ($\sigma ^{-}$, spin -1) luminescence. With
this quasi-resonant excitation, an energy splitting of 2.5 meV is clearly
seen between the two peaks in Fig.1a. Increasing the excitation 
density, both 
a broadening of the lines, which increases from 12 to 18 meV,
and a strong enhancement of the splitting is observed. The splitting is 
mostly due to the red shift of the $\sigma ^{-}$ polarized emission and 
exhibits marked time- and excitation-energy dependences.

The time dependence of the polarized (solid points) and unpolarized (open
points) photoluminescence is shown in Fig. 2 for two different excitation
energies and an initial carrier density of $5\times 10^{10}cm^{-2}$. At short
times, the splitting amounts to 4 meV exciting at 1.631 eV, below the
light-hole (lh) exciton (diamonds). However, if the excitation is moved at
energies above the lh it becomes only 1.2 meV (1.681 eV, circles). We have
found that the splitting is strongly correlated with the degree of
polarization (P) of the luminescence. Time-resolved measurements on the same
sample have shown that P at t = 10 ps amounts to 80$\%$ and 20$\%$ at 1.631
eV and 1.681 eV, respectively. \cite{Munoz_95} The behavior
of the peak positions of the PL with time, seen in the figure, is common for
all excitation energies: the polarized (unpolarized) emission shifts
towards lower (higher) energies with increasing time until
both emission bands merge at $\sim 150ps$.

Figure 3 shows the dependence of the energy positions of the luminescence
on the initial carrier density. The numbers in the abscissa have been
estimated from the absorption coefficient of the sample, the power density,
the area in which the laser beam is focused and the losses in the windows of
the cryostat, and are subjected to substantial uncertainties. Under the conditions
presented in the figure, 12 ps after the excitation at 1.631 eV, the $\sigma
^{+}$ emission remains practically constant, while the $\sigma ^{-}$
luminescence red shifts with increasing carrier density up to $~\sim
7\times 10^{10}cm^{-2}$. For higher carrier densities the energy of the
unpolarized emission stays constant. The solid lines correspond to the
calculation of the excitonic energies according to the method described in
Ref. \cite{Joaq_96}, which takes into account inter-excitonic exchange
interaction and screening. The changes in the energies of the
interacting +1 and -1 excitons with respect to the energy of a single
exciton are written analytically as:

\begin{eqnarray}
E^{\pm }=2(n^{\pm }+fn^{\mp })(I_1-I_2)
-\frac{0.82\pi e^2 na}{\epsilon }
\end{eqnarray}

where $n$ is the total density of excitons, $n^{\pm }$ are
those of $\pm 1$ excitons, 
$\epsilon $ the dielectric constant and $a$ the three-dimensional
Bohr radius. 
$I_2$ describes a "self-energy" correction that weakens electron-electron 
and hole-hole repulsion. $I_1$ is a "vertex" correction
that, due to the Pauli exclusion, reduces the inter-excitonic 
electron-hole attraction. 
The term involving $f$ is a small coupling between $\pm 1$ excitons, 
essentially due to valence band mixing. Since a $+3/2 (-3/2)$ 
hole has a small 
coupling with a $-1/2 (+1/2)$ hole, 
a hh $+1 (-1)$ exciton has weight mainly 
on $-1/2(+1/2)$ electrons but also some weight on $+1/2(-1/2)$ electrons. 
The last term in Eq. (1) is a screening correction using the 
random phase approximation.

Using the expressions for I$_1$ and I$_2$ given in the Appendix of 
Ref. \cite {Joaq_96}, with $\epsilon =13$ for GaAs and neglecting the 
small $f$ term, Eq. (1) can be re-written as:
\begin{eqnarray}
\ E^{\pm }(eV)=2.214\times 10^{-16}a(\AA )\times \nonumber \\
\times \left[ 1.515n^{\pm }(cm^{-2})-0.41\pi n(cm^{-2})\right] 
\end{eqnarray}

In the lines of figure 3, the energy of a single exciton, corresponding to
the limit $n\rightarrow 0$, has been taken as the experimental energy of the
+1 exciton at the lowest carrier density used in the experiments
($~\sim 6.5\times 10^9cm^{-2}$). 
A three-dimensional excitonic Bohr radius of 150 \AA\ and
an initial degree of polarization P= 80$\%$, corresponding to $n^{+}$= 0.9$n$
and $n^{-}$= 0.1$n$, have been used to compute the curves. The assumption
that P does not depend on $n$ is corroborated by previous 
TRPL experiments.\cite {Munoz_95} The results are 
plotted up to $9.5\times 10^{10}cm^{-2}$, where a
saturation of the splitting of the photoluminescence is observed.

The theory obtains a very good agreement with the experiments for $\sigma
^{+}$ emission and agrees qualitatively with the dependence of the $\sigma
^{-}$ luminescence. The flatness of the calculated position for the
polarized ($\sigma ^{+}$) PL arises from an almost perfect cancellation of
the "self-energy", the vertex correction and the 
screening in Eq. (2) for $E^{+}$. 
On the other hand, the theory predicts a less pronounced red-shift
of the unpolarized ($\sigma ^{-}$) PL with increasing carrier density than
that observed experimentally, and does not reproduce the saturation of the
shift. These discrepancies between the experiments and the calculations can
originate from different sources: i) the theory depends on excitonic
density, while the experiments are plotted against carrier density, which is
not necessarily the same as the excitonic one, especially at high densities
when the number of created excitons saturates; ii) the densities have
considerable uncertainties in their estimation; iii) the theory assumes free
excitons with zero kinetic energy, while the excitons are actually weakly
bound, \cite{Munoz_95} and have some excess energy because they are not
strictly created under resonant conditions; iv) the theory considers a 
strictly 
two-dimensional system while the actual one has a width in the growth 
direction. This last approximation 
distorts the relative importance of exchange 
versus Hartree interaction. 

The interdependence between the magnitude, and also the sign, 
of the splitting
and P is further demonstrated in the inset of Fig. 3, which depicts the 
shift of the PL peaks versus initial carrier density, 
exciting at the lh exciton with t = 12 ps.
In this case, we observe that the 
unpolarized emission (open triangles) lies
at slightly higher energies than the polarized one (closed triangles), in
concordance with the small, but not negligible, value of P, which amounts to
-10$\%$. \cite{Munoz_95} The negative value of P corresponds to the fact
that exciting at the lh-exciton energy with $\sigma ^{+}$ polarization, 
after relaxation of -1/2 lh to 3/2 hh, the
population of -1 hh-excitons is larger than that of +1 hh-excitons. 
The adequacy of the theory to explain the experimental results is also
confirmed by the results obtained with Eq.(2) for P=-10$\%$, 
which are depicted in the inset as 
solid ($\sigma ^{+}$) and open ($\sigma ^{-}$)
squares
for an excitonic density of $9.5\times 10^{10}cm^{-2}$. 
Although the theory gives again a less
pronounced dependence (not shown) on excitation power than 
the experimental one displayed in
the inset, it obtains for P=-10$\%$   
that both excitons redshift.
Furthermore, the calculated shift of the +1 excitons 
and the splitting amount to
2 meV and -0.4 meV, respectively, in very good agreement with the
experimental values.

The theory predicts (neglecting the second term in Eq. (1))
that the splitting between +1 and -1 excitons 
is proportional to the difference between
spin-up ($n^{+}$) and spin-down ($n^{-}$) populations 
and therefore
proportional to the degree of polarization 
(P=($n^{+}$-$n^{-}$)/($n^{+}$+$n^{-}$)). 
This dependence is also borne
out from the experiments, as shown
in Fig. 4: the time dependence of the splitting (solid points) and of the
degree of polarization (open diamonds), for a carrier density of $
5\times 10^{10}cm^{-2}$, are strongly correlated and both show a monoexponential
decay with a time constant of 40 ps. When the $n^{+}$ and $n^{-}$
populations become comparable, and therefore $P\rightarrow 0$, the splitting
vanishes as a consequence of the convergence of the +1 and -1 excitons
towards the same energy, as predicted by Eqs. (1) and (2). The reduction of P
with time arises from spin-flip processes of the excitons, which are believed to be
driven by exchange interaction between the electron and hole
composing the exciton. \cite{Damen_91,Maialle_93} This intra-excitonic
exchange interaction does not break the symmetry between spin +1 and spin -1
excitons and has a very weak influence in the exciton energy levels, but it
is responsible for the time decay of the splitting.

Finally, we would like to mention that the experiments corroborate the
theoretical predictions concerning the relative strength of the 
electron-hole vertex (I$_1$) 
and of the electron(hole)-electron(hole) (I$_2$) "self-energy" 
corrections: the splitting grows with increasing initial carrier
density corresponding to $I_1/I_2=6.28/4.76>1$
(see Eqs. (A4) and (A12) in Ref. \cite{Joaq_96}). The possible
reversal of this inequality by an external perturbation could have important
consequences in the state of polarization of the excitonic gas. The
variation of the total energy with the degree of polarization P can be
written as:

\begin{equation}
\frac{\partial E_T}{\partial P}=\frac{2e^2n^2}{\epsilon a} (I_1-I_2) (1-f) P
\end{equation}

Taking into account that $f$ $<<$ 1, if I$_1$ $>$ I$_2$, Eq. (3) predicts 
that the system prefers to have zero
polarization because there it attains an energy minimum. However, if an external
perturbation, such as strain, electric field, etc., could cause
that I$_2$ $>$ I$_1$ then the excitonic gas would prefer to be polarized. \cite
{Joaq_ber}

In summary, we have shown that the spin splitting observed in a polarized
two-dimensional exciton gas originates from the inter-excitonic 
interaction among electrons and holes, forming the excitons.
The magnitude of the splitting depends on the process of excitonic
formation, it grows with increasing excitonic density and is firmly 
correlated with the degree of polarization of the system. Its time evolution
is determined by the aditional intra-excitonic exchange interaction.

\section{Acknowledgments}

This research was supported in part by the Comisi\'on Interministerial de
Ciencia y Tecnolog\'{\i}a of Spain under contract MAT 94-0982-C02-01 and by
the Commission of European Communities under contract Ultrafast
CHRX-CT93-0133.

\begin{figure}[tbp]
\caption{Low temperature, 8K, 
time-resolved PL spectra of a 77$\AA$-wide GaAs QW taken 10 ps 
after the excitation with $\sigma ^{+}$ polarized light at 1.625 eV. The 
solid (open) points depict the $\sigma ^{+}$($\sigma ^{-}$) emission. Initial
carrier density: a) $4\times 10^{10}cm^{-2}$, b) $1.5\times
10^{11}cm^{-2}$. The arrows
indicate the blue(red) shift of the 
$\sigma ^{+}$($\sigma ^{-}$) luminescence.}
\label{fig1}
\end{figure}

\begin{figure}[tbp]
\caption{Time evolution of the excitonic energies 
for $\sigma ^{+}$ (solid points) and $\sigma ^{-}$
(open points) emission at two 
excitation energies: 1.631 eV ($\diamond$) and
1.681 eV ($\circ$), 
for an initial carrier density of $5\times 10^{10}cm^{-2}$. 
The exciting light was $\sigma ^{+}$ polarized.} 
\label{fig2}
\end{figure}

\begin{figure}[tbp]
\caption{Energies of the polarized ($\sigma ^{+}$, solid points) and 
unpolarized($\sigma ^{-}$, open points) luminescence as a function of
carrier density. The positions are taken 12 ps after the excitation
at 1.631 eV. The lines represent the results of Eq. (2). 
The inset shows the dependence on carrier density of the split
luminescence at 10 ps after the excitation at the light-hole 
exciton energy.
The squares show the predictions of the theoretical model and 
the dashed lines are a guide to the eye.}
\label{fig3}
\end{figure}

\begin{figure}[tbp]
\caption{Time evolution of the PL splitting $(\bullet )$ and polarization
($\diamond$) for an initial carrier density of $5\times 10^{10}cm^{-2}$ exciting
at 1.621 eV. The dashed line depicts the best fit to a monoexponential
decay with a time constant of 41 ps.}
\label{fig4}
\end{figure}


\begin{references}
\bibitem{Damen_91}T.C. Damen, L. Vi\~na, J.E. Cunningham, J. Shah, and
L.J. Sham, Phys. Rev. Lett. {\bf 67}, 3432 (1991).

\bibitem{Stark_92}J.B. Stark, W.H. Knox, and D.S. Chemla, Phys. Rev. B 
{\bf 46}, 7919 (1992).

\bibitem{Dareys_93}B. Dareys, X. Marie, T. Amand, J. Barrau, Y. Shekun, I.
Razdobreev, and R. Planel, Superlatt. and Microstruct. {\bf 13}, 353 (1993).

\bibitem{Amand_94}T. Amand, X. Marie, B. Baylac, B. Dareys, J. Barrau, M.
Brousseau, R. Planel, and D.J. Dunstan, Phys. Lett. A {\bf 193}, 105 (1994).

\bibitem{Vinattieri_93}A. Vinattieri, J. Shah, T.C. Damen, K.W. Goosen,
L.N. Pfeiffer, M.Z. Maialle, and L.J. Sham, Appl. Phys. Lett. {\bf 63}, 3164
(1993).

\bibitem{Munoz_95}L.Mu\~noz, E.P\'erez, L. Vi\~na, and K. Ploog, Phys.
Rev. B {\bf 51}, 4247 (1995).

\bibitem{Bastard_92}G. Bastard and R. Ferreira, Surf. Sci. {\bf 267}, 335
(1992).

\bibitem{Maialle_93}M.Z. Maialle, E.A. de Andrada e Silva, and L.J. Sham,
Phys. Rev. B {\bf 47}, 15576 (1993).

\bibitem{Joaq_96}J. Fern\'andez-Rossier, C. Tejedor, L. Mu\~noz, and L.
Vi\~na, Phys. Rev. B {\bf 53}, ???? (1996).

\bibitem{com_1}In this work we denote the third component of the exciton
angular moment with the word {\em spin}.

\bibitem{Munoz_95b}L.Mu\~noz, E.P\'erez, L. Vi\~na, J.Fern\'andez-Rossier,
C. Tejedor and K. Ploog, Solid St. Elec. {\bf 40}, 755 (1996). 

\bibitem{Peyghambarian_84}N. Peyghambarian, H.M. Gibbs, J.L. Jewell, A.
Antonetti, A. Migus, D. Hulin, and A. Myssyrowicz, 
Phys. Rev. Lett. {\bf 53}, 2433 (1984).

\bibitem{Hulin_86}D. Hulin, A. Myssyrowicz, A. Antonetti, A. Migus, W.T.
Masselink, H. Morkoc, H.M. Gibbs, and N. Peyghambarian, 
Phys. Rev. B {\bf 33}, 4389 (1986).

\bibitem{Rink_85}S. Schmitt-Rink, D.S. Chemla and D.A.B. Miller, Phys.
Rev. B {\bf 32}, 6601 (1985).

\bibitem{Haug_94}H. Haug and S. Schmitt-Rink, Prog. Quant. Electr. {\bf 9},
3 (1984).

\bibitem{Joaq_ber}J. Fern\'andez-Rossier and C. Tejedor,
in Proceedings of the 23rd International Conference on the 
Physics of Semiconductors, edited by M. Scheffler and R. Zimmermann 
(World Scientific, Singapore, 1996), p.2463.

\end{references}
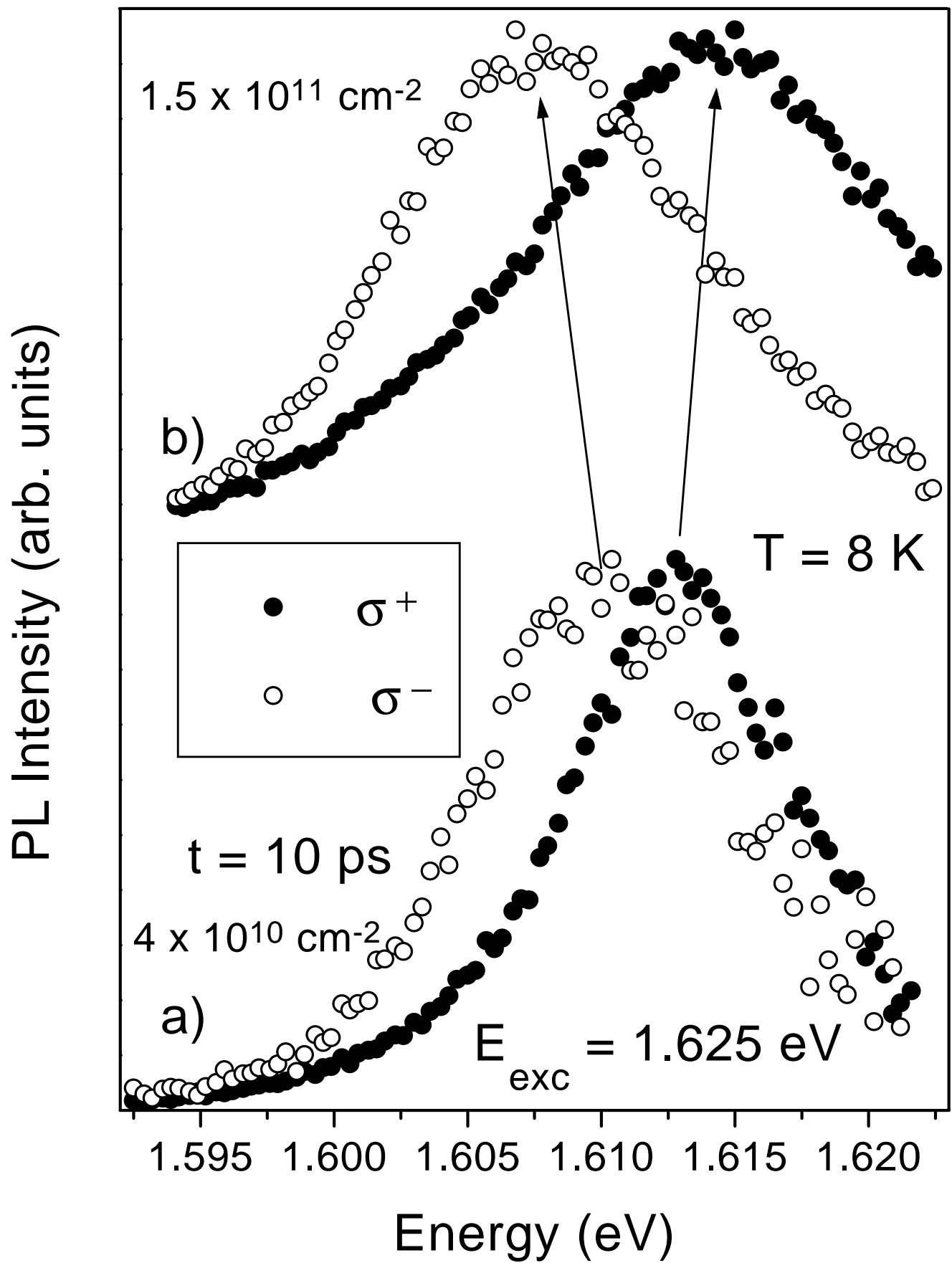
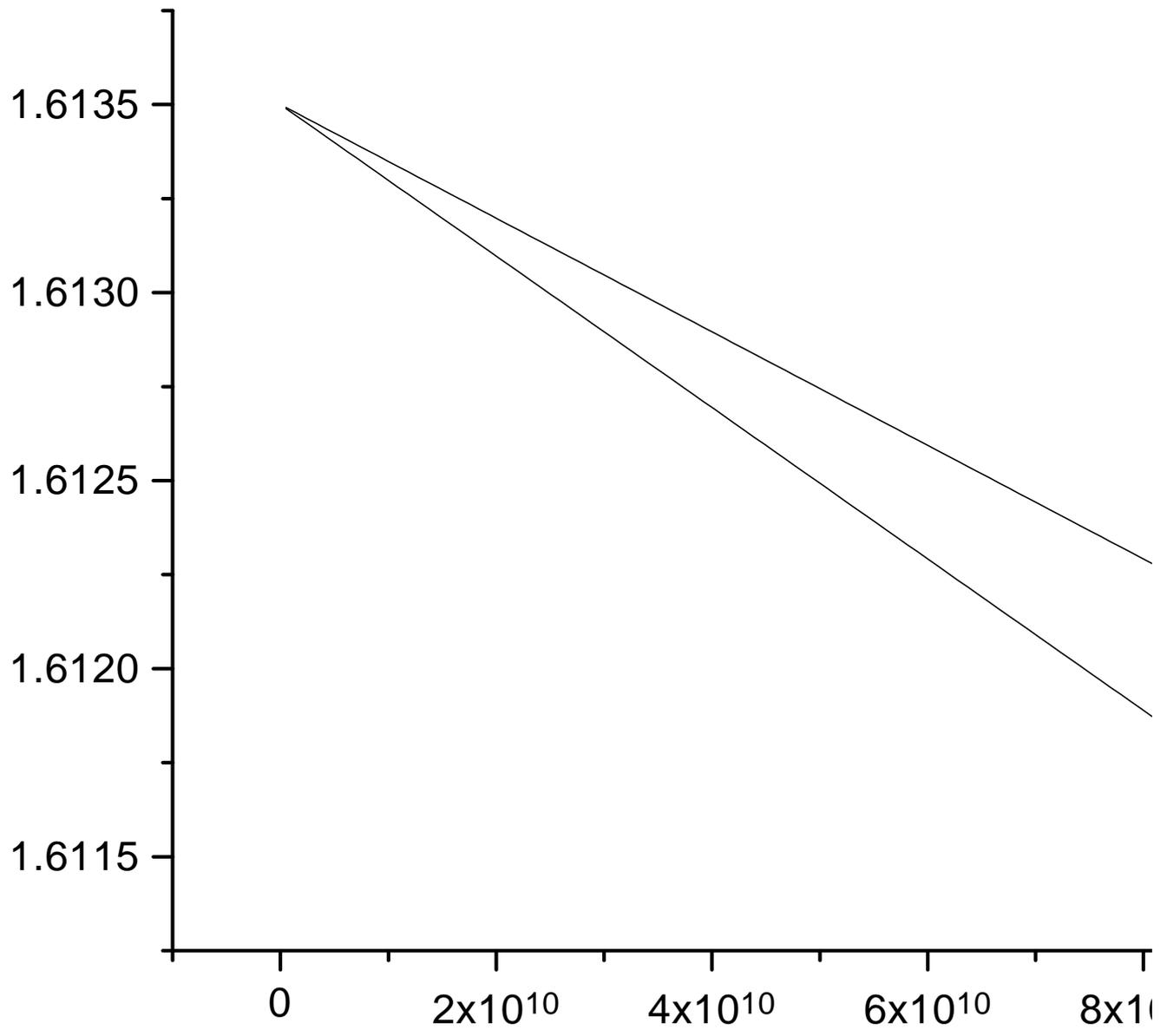
\end{document}